# Correlates of repeat abortions and their spacing: Evidence from registry data in Spain

Catia Nicodemo[1,2,3], Sonia Oreffice[3,4,5], Climent Quintana-Domeque[3,4,5,*]

May 2023[**]

## Abstract

Using administrative data on all induced abortions recorded in Spain in 2019, we analyze the characteristics of women undergoing repeat abortions and the spacing between these procedures. Our findings indicate that compared to women experiencing their first abortion, those who undergo repeat abortions are more likely to have lower education levels, have dependent children, live alone, or be foreign-born, with a non-monotonic relationship with age. We also report that being less educated, not employed, having dependent children, or being foreign-born are all strongly related to a higher number of repeat abortions. Lastly, we find that being less educated, foreign-born, or not employed is correlated with a shorter time interval between the last two abortions.

[1] Nuffield Department of Primary Care, Health Sciences, Medical Sciences Division, University of Oxford, Oxford, United Kingdom.
[2] Department of Economics, University of Verona, Verona, Italy.
[3] IZA, Bonn, Germany.
[4] Department of Economics, Business School, University of Exeter, Exeter, United Kingdom.
[5] HCEO, Department of Economics, University of Chicago, Chicago, USA.
[*] c.quintana-domeque@exeter.ac.uk
[**] This is a revised version of a previous preprint submitted on 10 August 2022 with the title "Risk factors for repeat abortions and their spacing: Evidence from registry data in Spain."

# Introduction

Abortion provides women with the means to terminate unintended pregnancies under medically warranted circumstances, instances of contraceptive failure or unavailability, or after encountering physical or psychological violence. On June 24, 2022, the US Supreme Court ruled in Dobbs v. Jackson Women's Health Organization (https://www.supremecourt.gov/opinions/21pdf/19-1392_6j37.pdf) to overturn the constitutional right of American women to access abortion. This ruling impairs women's access to abortion and family planning services, especially affecting the poor and vulnerable (a review of changes in the historical policy environment in the United States that serves as the foundation of the empirical literature that estimates the causal effects of access to contraception and abortion has been offered [1]). This significant social change and opposition to women's health rights affect millions of women worldwide, underscoring the need for research on abortion. We must better understand the women most at risk and to develop health policies that serve their needs.

Access to abortion, or lack thereof, not only affects maternal health [2] but also profoundly influences women's lives. It impacts their educational attainment, labor force participation, overall earnings, and balance of power within a couple [3–7]. It is essential to distinguish between contraception, which aims to prevent pregnancy, and abortion, which addresses an established unintended pregnancy.

Repeat, or multiple, abortions occur when a woman undergoes more than one abortion in her lifetime due to a variety of factors [8]. Reducing the need for multiple abortions involves understanding the reasons behind repeat abortions and providing comprehensive reproductive healthcare, including effective contraception and support services.

In this study, we examine the factors that correlate with the probability of experiencing a repeat abortion, the number of repeat abortions, and the interval between the last and previous abortions in Spain for 2019. We use administrative data from the Ministry of Health, Social Services, and Equality. This comprehensive dataset includes all recorded induced abortions in 2019 throughout the country, the number of repeat abortions, the date of the previous abortion (if any), and a rich set of women's demographic and socioeconomic characteristics.

In Spain, women can choose to have an induced abortion for any reason until the 14$^{th}$ week, and for medical reasons between the 14$^{th}$ and the 21$^{st}$ week (Law of Sexual and Reproductive Health 2010: https://www.boe.es/eli/es/lo/2010/03/03/2/con). In 2019, the abortion rate in Spain was 11.53 per 1,000 women aged 15-44 (https://www.sanidad.gob.es/en/profesionales/saludPublica/prevPromocion/embarazo/docs/IVE_2019.pdf), just above the average for the period 2010-2018 (11.21). The total number of abortions was 99,149, of which 35.87% were repeat abortions (https://www.sanidad.gob.es/en/profesionales/saludPublica/prevPromocion/embarazo/docs/IVE_2019.pdf). Of repeat abortions, 79% were surgical, compared to 74% of first-time abortions. The percentage of repeat abortions in Spain is very similar to that for the Netherlands (33%), lies within the ranges of estimates for other European countries (30-38%), but is much below that estimated for the US (47%) [9]. Thus, repeat abortions are still a widespread phenomenon.

We find that women who undergo a repeat abortion, compared to those experiencing their first abortion, tend to be less educated and more likely to have dependent children, live alone, or be foreign-born. Additionally, we observe a non-monotonic relationship with age. Focusing on the actual number of abortions, we show that low education, not being employed, having dependent children, or being foreign-born are strongly related to a higher number of repeat abortions. Lastly, we establish that the socioeconomic factors associated with a shorter time interval (between a woman's latest 2019 abortion and her previous one) include low education, not being employed, and being foreign-born.

There is limited research on the characteristics of women who undergo repeat abortions, and studies vary widely in terms of data availability, from survey data of one hospital in the US [10] to registration data of abortion clinics in the Netherlands responsible for about two thirds of abortion procedures in 2010 [9]. However, several consistent findings have been documented. Repeat abortions are more likely among women who have been using some method of birth control [11,12], among unmarried women [12,13], among women with more children [11,14,15,16], and among low-educated women [15,16]. The findings regarding age are quite mixed: while some studies found older women have a higher risk [10,11,14,17], others found younger women have a higher risk [15,18].

Given our findings, healthcare services and policymakers in Spain should target family planning and contraception services more effectively at the time of a woman's first abortion, especially tailored for high-risk groups: low-educated, those without employment, or foreign-born.

## Materials and methods

This cross-sectional study examines women in Spain who sought abortions in 2019. In Spain, the Law of Sexual and Reproductive Health passed in 2010 allows any adult woman to choose an induced abortion within the first 14 weeks of gestation (https://www.boe.es/eli/es/lo/2010/03/03/2/con). After the 14$^{th}$ week and up to the 21$^{st}$ week, induced abortions can be performed for medical reasons.

We utilize a comprehensive dataset of all recorded abortions in Spain (N=99,149) from the Ministry of Health, Social Services, and Equality. The Spanish reporting system for abortions provides reliable and complete data for studying the incidence and correlates of repeat abortions. Each legal induced abortion is logged into the system of the accredited hospital/clinic performing it and periodically submitted (validated and encrypted) to the Ministry of Health, Social Services, and Equality in Spain via an electronic database (https://www.sanidad.gob.es/profesionales/saludPublica/prevPromocion/embarazo/home.htm). Instructions on how to access the data are available online (https://www.sanidad.gob.es/en/profesionales/saludPublica/prevPromocion/embarazo/docs/Informe_MetodologicoIVE.pdf). The administrative data are anonymized, and institutional review board was not required. Access to the database with the universe of induced abortions was obtained from the Spanish Ministry of Health in June 2021.

Our aim is to assess the characteristics associated with a history of at least one previous abortion, the number of previous abortions, and the spacing of abortions for women undergoing a repeat abortion in 2019. Our main outcomes of interest are: (a) a repeat abortion binary indicator (=0 if no previous abortions, =1 if previous abortions); (b) a count number of previous abortions measure (number of self-reported previous abortions); and (c) the number of months elapsed between the current abortion and the previous one among women who had (at least) a previous abortion (computed using the information on the month and year of the previous and current abortion). Our main explanatory variables are: age of the woman in 5-year brackets (from 15-19 to 40-44), education (primary or less, secondary, university), employment status (employed vs. not employed), foreign born (=0 if born in Spain, =1 else), living arrangement indicators (alone, in a couple, with family members, with others), children (=0 if no dependent children, =1 if dependent children), use of contraceptive methods (=0 if the woman does not regularly use contraceptive methods, =1 if the woman regularly uses them), and an indicator of whether the abortion was publicly funded (= 0 if no, = 1 if yes). Moreover, all regressions include woman's province of residence (50 provinces). We also consider information on whether there were serious health reasons for the abortion or if it was an elective procedure, as well as the number of weeks of gestation to exclude rare abortions performed in the third trimester of pregnancy. The survey questionnaire defines the reason behind the interruption of the pregnancy based on the specific law of sexual and reproductive health and elective interruption of pregnancy in Spain (https://www.boe.es/eli/es/lo/2010/03/03/2/con).

Analyses were conducted with Stata statistical software version 17. The code will be publicly available from the Harvard Dataverse repository upon publication.

## Results

Our final sample size consists of 97,921 abortions across 50 provinces, focusing on the age group of 15-44, which accounts for 98.98% of all abortions (N=98,147). We exclude 149 rare abortions performed in the third trimester of pregnancy (25 weeks of gestation or later), 14 abortions in the autonomous city of Ceuta, and 63 abortions in the autonomous city of Melilla (both located on the African continent).

Out of the 97,921 abortions, 36.04% were repeat abortions. Among those women who had previously undergone an abortion, the average (median) number of previous abortions was 1.57 (1): 65.67% had one previous abortion, 21.72% had two previous abortions, 7.53% had three previous abortions, and the remaining 5.08% had four or more previous abortions. Additionally, the average time interval between the last abortion in 2019 and the previous one was approximately 56.5 months (slightly more than 4.5 years) – 56.38 months, or 56.59 months if we focus on an interval of at least 3 months between the abortion in 2019 and the previous one.

Of the 97,921 abortions, 91.08% were elective (i.e., requested by the woman), while the remaining 8.92% were due to medical reasons (including serious risk to the life or health of the pregnant woman, risk of severe fetal anomalies, and fetal anomalies incompatible with life or very serious untreatable/incurable illness).

# Single vs. repeat abortions

In Table 1, we present the distribution of characteristics of women who sought an abortion (first vs. repeat) in 2019. This provides a univariate analysis of the relationship between the likelihood of repeat abortion vs. first abortion and each demographic characteristic, one at a time. All characteristics, except for the reported use of contraceptive methods, appear not to be independent of the type of abortion (first vs. repeat), as judged by the p-value of the Chi-square test.

**Table 1. Distribution of women characteristics by first vs. repeat abortion.**

| Characteristic | First abortion (n) | Repeat abortion (n) | p-value |
|---|---|---|---|
| Age (y) (N=97,921) | | | 0.000 |
| 15-19 | 8,650 | 1,369 | |
| 20-24 | 14,441 | 6,470 | |
| 25-29 | 13,148 | 8,685 | |
| 30-34 | 11,544 | 8,975 | |
| 35-39 | 10,174 | 7,103 | |
| 40-44 | 4,671 | 2,871 | |
| Education (N=96,490) | | | 0.000 |
| Primary or less | 8,619 | 6,597 | |
| Secondary | 39,635 | 24,273 | |
| University | 13,389 | 3,977 | |
| Employed (N=95,827) | | | 0.002 |
| No | 23,896 | 13,159 | |
| Yes | 37,330 | 21,442 | |
| Foreign born (N=96,751) | | | 0.000 |
| No | 40,189 | 19,865 | |
| Yes | 21,540 | 15,157 | |
| Living arrangement (N=96,075) | | | 0.000 |
| Alone | 12,620 | 7,543 | |
| Living in couple | 27,739 | 18,680 | |
| Living with parents / family | 18,622 | 7,409 | |
| Living with others | 2,473 | 989 | |
| Dependent children (N=96,007) | | | 0.000 |
| No | 34,189 | 11,162 | |
| Yes | 27,047 | 23,609 | |
| Trimester of pregnancy (N=97921) | | | 0.000 |
| First | 56,045 | 32,675 | |
| Second | 6,583 | 2,618 | |
| Contraceptive methods (N=85,679) | | | 0.605 |
| No | 23,434 | 13,550 | |
| Yes | 30,938 | 17,757 | |
| Publicly funded abortion (N=97,921) | | | 0.000 |
| No | 16,598 | 7,549 | |
| Yes | 46,030 | 27,744 | |
| Elective (N=97,921) | | | 0.000 |
| No | 6,414 | 2,318 | |
| Yes | 56,214 | 32,975 | |

Notes: p-value from a Chi-square test of independence between rows and columns for each variable. First trimester of pregnancy (≤ 12 weeks of gestation). Second trimester of pregnancy (13-24 weeks of gestation).

In Table 2, we report the average differences between those who had a repeat abortion in 2019 and those who had a first abortion in 2019. Women who had a repeat abortion were, on average, 2 years older (mean = 30.19, SD = 6.43) than those who underwent their first abortion (mean = 28.14, SD = 7.38). They were 11 percentage points (pp) less likely to have a university degree (11% vs. 22%), 8 pp more likely to be foreign-born (43% vs. 35%), 9 pp more likely to be living in a couple (54% vs. 45%), and 24 pp more likely to have dependent children (68% vs. 44%). They were also 4 pp more likely to have their abortion in the 1st trimester (≤12 weeks of gestation) of pregnancy (93% vs. 89%), 6 pp more likely to have their abortion publicly funded (79% vs. 73%), and 3 pp more likely to have an elective abortion (requested the abortion for non-medical reasons, 93% vs. 90%). We do not find differences in the likelihood of the mother regularly using contraceptive methods (57% in both cases), and there is a 1 pp difference in the likelihood of being employed (62% among those who had a repeat abortion vs. 61% among those who had their first abortion).

Table 2. Comparison of average characteristics between first abortion and repeat abortion.

| Characteristic | First Abortion | | Repeat Abortion | | | |
| --- | --- | --- | --- | --- | --- | --- |
| | Mean | N | Mean | N | Diff | p-value |
| Age | 28.14 | 62,628 | 30.19 | 35,293 | 2.05 | 0.000 |
| University | 0.22 | 61,643 | 0.11 | 34,847 | -0.11 | 0.000 |
| Employed | 0.61 | 61,226 | 0.62 | 34,601 | 0.01 | 0.002 |
| Foreign born | 0.35 | 61,729 | 0.43 | 35,022 | 0.08 | 0.000 |
| Living alone | 0.21 | 61,454 | 0.22 | 34,621 | 0.01 | 0.000 |
| Living in couple | 0.45 | 61,454 | 0.54 | 34,621 | 0.09 | 0.000 |
| Living with parents / family | 0.30 | 61,454 | 0.21 | 34,621 | -0.09 | 0.000 |
| Living with others | 0.04 | 61,454 | 0.03 | 34,621 | -0.01 | 0.000 |
| Dependent children | 0.44 | 61,236 | 0.68 | 34,771 | 0.24 | 0.000 |
| Trimester 1 (≤ 12 weeks) | 0.89 | 62,628 | 0.93 | 35,293 | 0.04 | 0.000 |
| Use contraception methods | 0.57 | 54,372 | 0.57 | 31,307 | 0.00 | 0.605 |
| Publicly funded | 0.73 | 62,628 | 0.79 | 35,293 | 0.06 | 0.000 |
| Elective | 0.90 | 62,628 | 0.93 | 35,293 | 0.03 | 0.000 |

Note: p-value from the difference in means is obtained as the p-value that the slope coefficient from a linear regression of the variable (characteristic) on a constant and a repeat abortion indicator (=0 if first abortion, =1 if repeat abortion) is zero, using robust standard errors.

In Table 3, we investigate the predictors of repeat abortions among women who had an abortion in 2019 in a multivariate setting. While we discuss estimates based on ordinary least squares regressions here, **S1 Table in S1 Appendix** contains logit estimates and displays the findings in odds ratios format. We run our multivariate regression for three different samples: (1) full sample, (2) adult sample age 18-44 (i.e., full sample excluding minors), and (3) adult sample age 18-44 excluding abortions due to health reasons.

**Table 3. Correlates of repeat abortion among women who had an abortion in 2019.**
**Dependent variable = 1 if repeat abortion, = 0 no previous abortion**
**Linear probability model: OLS regression estimates (95% Confidence intervals)**

|  | (1) | (2) | (3) |
|---|---|---|---|
| Age |  |  |  |
| 15-19 | -0.197*** | -0.166*** | -0.175*** |
|  | (-0.213 , -0.180) | (-0.184 , -0.149) | (-0.194 , -0.156) |
| 20-24 | -0.029*** | -0.031*** | -0.042*** |
|  | (-0.044 , -0.014) | (-0.046 , -0.016) | (-0.058 , -0.026) |
| 25-29 | 0.041*** | 0.040*** | 0.031*** |
|  | (0.027 , 0.055) | (0.026 , 0.054) | (0.016 , 0.046) |
| 30-34 | 0.053*** | 0.052*** | 0.047*** |
|  | (0.039 , 0.067) | (0.038 , 0.066) | (0.032 , 0.062) |
| 35-39 | 0.026*** | 0.025*** | 0.022*** |
|  | (0.011 , 0.040) | (0.011 , 0.040) | (0.007 , 0.038) |
| Education |  |  |  |
| University | -0.212*** | -0.218*** | -0.214*** |
|  | (-0.224 , -0.200) | (-0.230 , -0.206) | (-0.227 , -0.201) |
| Secondary | -0.053*** | -0.060*** | -0.059*** |
|  | (-0.063 , -0.044) | (-0.070 , -0.049) | (-0.070 , -0.048) |
| Employed | -0.016*** | -0.019*** | -0.016*** |
|  | (-0.024 , -0.009) | (-0.026 , -0.012) | (-0.024 , -0.009) |
| Living arrangements |  |  |  |
| Living alone | 0.080*** | 0.081*** | 0.081*** |
|  | (0.062 , 0.097) | (0.063 , 0.099) | (0.063 , 0.100) |
| Living in a couple | 0.045*** | 0.046*** | 0.055*** |
|  | (0.028 , 0.061) | (0.029 , 0.062) | (0.038 , 0.072) |
| Living with relatives | 0.041*** | 0.045*** | 0.046*** |
|  | (0.025 , 0.058) | (0.028 , 0.062) | (0.029 , 0.064) |
| Dependent children | 0.138*** | 0.137*** | 0.136*** |
|  | (0.130 , 0.146) | (0.129 , 0.145) | (0.127 , 0.145) |
| Foreign born | 0.047*** | 0.046*** | 0.042*** |
|  | (0.040 , 0.054) | (0.039 , 0.054) | (0.034 , 0.050) |
| Use contraceptive methods | 0.017*** | 0.018*** | 0.008** |
|  | (0.011 , 0.024) | (0.011 , 0.025) | (0.001 , 0.015) |
| Publicly funded abortion | 0.019*** | 0.019*** | 0.025*** |
|  | (0.011 , 0.028) | (0.010 , 0.027) | (0.015 , 0.034) |
| Observations | 81,168 | 78,129 | 72,052 |
| R-squared | 0.095 | 0.084 | 0.085 |
| Adults only? | No | Yes | Yes |
| Elective only? | No | No | Yes |

Note: Reference category: women aged 40-44, with primary education or less, not employed, living with others, with no children in charge, born in Spain, not using contraceptive methods, whose abortion has not been publicly funded, and whose province of residence is Álava.
95% confidence intervals based on robust standard errors in parentheses.
*** p<0.01, ** p<0.05, * p<0.1

Focusing on the estimates in column (1), we find evidence of a non-monotonic relationship between age and the likelihood of a repeat abortion. The risk of repeat abortion increases with age from 15-19 (point estimate: -0.197, 95% CI: [-0.213, -0.180]) to 30-34 (0.053, [0.039, 0.067]), but then it decreases from 30-34 to 40-44 (reference category: 0). Women who have a university degree are about 20 percentage points less likely to have had an abortion previously than those with primary education or less (-0.212, [-0.224, -0.200]). The gap among women with secondary education and primary education or less is 5 percentage points (-0.053, [-0.063, -0.044]). Employed women who had an abortion in 2019 were about 1.6 percentage points less likely to have had a prior abortion (-0.016, [-0.024, -0.009]). Women who live alone (0.080, [0.062, 0.097]), live in a couple (0.045, [0.028, 0.061]), or live with relatives (0.041, [0.025, 0.058]) are more likely to have a repeat abortion than those who live with others. Having dependent children is associated with an increase in the likelihood of having another abortion of 14 percentage points (0.138, [0.130, 0.146]). Foreign-born women are about 4 percentage points more likely to have a repeat abortion than those born in Spain (0.047, [0.040, 0.054]). Finally, women who report using contraceptive methods regularly are more likely to have a repeat abortion (0.017, [0.011, 0.024]), and the likelihood of repeat abortion is also higher if the abortion is publicly funded (0.019, [0.011, 0.028]).

The estimates displayed in columns (2) and (3) are qualitatively identical and quantitatively very similar to those in column (1). Perhaps the two main differences are found, first, when looking at the age coefficients and, second, when looking at the coefficient on the woman reporting using contraceptive methods regularly. The finding on the age coefficients is not surprising since the indicator variable 15-19 contains only 18- and 19-year-olds in columns (2) and (3). The discrepancy when looking at the coefficient on the woman reporting using contraceptive methods regularly is likely to be driven by "collider bias", if both the type of abortion and the use of contraceptive methods are potential determinants of elective abortion behavior.

# Number of previous abortions

We now turn our attention to the determinants of the number of previous abortions among women who had an abortion in 2019. We estimate Poisson regressions to account for the count nature of the data. Table 4 reports estimates of incidence-rate ratios (IRRs) and their 95% confidence intervals.

Table 4. Correlates of number of previous abortions among women who had an abortion in 2019.
Dependent variable = previous number of abortions
Poisson regression: IRRs estimates (95% Confidence intervals)

|  |  | (1) | (2) | (3) |
|---|---|---|---|---|
| Age |  |  |  |  |
|  | 15-19 | 0.305*** | 0.386*** | 0.371*** |
|  |  | (0.281 , 0.330) | (0.356 , 0.419) | (0.341 , 0.404) |
|  | 20-24 | 0.796*** | 0.792*** | 0.764*** |
|  |  | (0.754 , 0.840) | (0.750 , 0.835) | (0.722 , 0.809) |
|  | 25-29 | 1.088*** | 1.085*** | 1.047* |
|  |  | (1.035 , 1.144) | (1.032 , 1.140) | (0.994 , 1.103) |
|  | 30-34 | 1.158*** | 1.156*** | 1.128*** |
|  |  | (1.102 , 1.216) | (1.101 , 1.215) | (1.072 , 1.188) |
|  | 35-39 | 1.067** | 1.066** | 1.043 |
|  |  | (1.014 , 1.122) | (1.014 , 1.122) | (0.989 , 1.099) |
| Education |  |  |  |  |
|  | University | 0.380*** | 0.377*** | 0.386*** |
|  |  | (0.362 , 0.398) | (0.360 , 0.395) | (0.368 , 0.405) |
|  | Secondary | 0.743*** | 0.738*** | 0.740*** |
|  |  | (0.720 , 0.766) | (0.716 , 0.762) | (0.717 , 0.764) |
| Employed |  | 0.878*** | 0.873*** | 0.882*** |
|  |  | (0.856 , 0.900) | (0.852 , 0.895) | (0.860 , 0.904) |
| Living arrangements |  |  |  |  |
|  | Living alone | 1.342*** | 1.344*** | 1.346*** |
|  |  | (1.248 , 1.443) | (1.250 , 1.445) | (1.249 , 1.452) |
|  | Living in a couple | 1.201*** | 1.202*** | 1.236*** |
|  |  | (1.120 , 1.289) | (1.121 , 1.290) | (1.149 , 1.330) |
|  | Living with relatives | 1.171*** | 1.180*** | 1.184*** |
|  |  | (1.089 , 1.258) | (1.097 , 1.268) | (1.099 , 1.276) |
| Dependent children |  | 1.633*** | 1.626*** | 1.604*** |
|  |  | (1.587 , 1.681) | (1.580 , 1.673) | (1.557 , 1.653) |
| Foreign born |  | 1.146*** | 1.143*** | 1.119*** |
|  |  | (1.117 , 1.175) | (1.114 , 1.172) | (1.090 , 1.148) |
| Use contraceptive methods |  | 0.968*** | 0.969*** | 0.939*** |
|  |  | (0.946 , 0.991) | (0.946 , 0.992) | (0.916 , 0.962) |
| Publicly funded abortion |  | 1.066*** | 1.064*** | 1.080*** |
|  |  | (1.033 , 1.100) | (1.031 , 1.098) | (1.044 - 1.118) |
| Observations |  | 81,168 | 78,129 | 72,052 |
| Adults only? |  | No | Yes | Yes |
| Elective only? |  | No | No | Yes |

Note: Reference category: women aged 40-44, with primary education or less, not employed, living with others, with no children in charge, born in Spain, not using contraceptive methods, whose abortion has not been publicly funded, and whose province of residence is Álava.
95% confidence intervals based on robust standard errors in parentheses.
*** p<0.01, ** p<0.05, * p<0.1

We begin with the description of the estimates in column (1). As when studying the determinants of repeat vs. first time abortion, the relationship is non-monotonic between age and number of previous abortions, increasing between the age group 15-19 (0.305, [0.281, 0.330]) to 30-34 (1.158, [1.102, 1.216]), and decreasing between the age group 30-34 to 40-44 (reference: 1), while holding the other variables constant. In terms of socioeconomic status, women with higher education and employed women are expected to have lower rates of repeat abortions: the rate among women with a university degree is less than half of that for women with primary education or less (0.380, [0.362, 0.398]), and the rate among employed women is also lower than that of women who do not have an employment (0.878, [0.856, 0.900]), that is, women who are classified as either a "Pensioner," "Student," "Unemployed," "Engaged in unpaid household work," or "Other." Women who live alone (1.342, [1.248, 1.443]), in a couple (1.201, [1.120, 1.289]), or with relatives (1.171, [1.089, 1.258]) exhibit a higher rate of repeat abortions than those living with others (reference: 1). Similarly, women with dependent children are expected to have a rate 1.6 times greater (1.633, [1.587, 1.681]) for number of repeat abortions than those without dependent children. Foreign-born women have an expected rate of repeat abortions 1.1 times greater (1.146, [1.117, 1.175]) than that of women born in Spain. The expected rate of repeat abortions among women who report using contraceptive methods regularly is lower than that of women who report not using them (0.968, [0.946, 0.991]). Finally, women relying on publicly funded abortions are expected to have a greater rate of number of repeat abortions (1.066, [1.033, 1.100]) than women whose abortions are not publicly funded.

Similar qualitative and quantitative results are found in columns (2) and (3). **S2 Table in S1 Appendix** contains Poisson regression estimates after excluding women reporting 6 or more previous abortions (0.43% of observations), with similar qualitative findings for all but one predictor: "women using contraceptive methods regularly".

## Spacing between abortions

Finally, in Table 5, we shift our attention to the determinants of the time span between abortions among women who had a repeat abortion in 2019. This is quite a unique feature of our data analysis: we have information on the actual month and year of the current as well as of the previous abortion. We can analyze which characteristics make it more likely for a woman who already had at least one abortion in the past to seek another abortion sooner rather than later. As before, we first focus on column (1).

**Table 5. Correlates of abortion spacing among women who had an abortion in 2019.**
**Dependent variable = months between current abortion and previous abortion**
**Linear regression model: OLS regression estimates (95% Confidence intervals)**

|  | (1) | (2) | (3) |
|---|---|---|---|
| Age |  |  |  |
| 15-19 | -72.917*** | -72.647*** | -71.848*** |
|  | (-76.472, -69.361) | (-76.335, -68.960) | (-75.560, -68.136) |
| 20-24 | -62.507*** | -62.160*** | -62.204*** |
|  | (-65.835, -59.179) | (-65.614, -58.707) | (-65.658, -58.749) |
| 25-29 | -46.505*** | -46.261*** | -46.284*** |
|  | (-49.815, -43.195) | (-49.701, -42.820) | (-49.725, -42.843) |
| 30-34 | -31.267*** | -31.087*** | -31.098*** |
|  | (-34.647, -27.887) | (-34.602, -27.573) | (-34.613, -27.584) |
| 35-39 | -16.705*** | -16.177*** | -16.181*** |
|  | (-20.282, -13.128) | (-19.899, -12.455) | (-19.903, -12.459) |
| Education |  |  |  |
| University | 8.123*** | 8.471*** | 8.512*** |
|  | (5.547, 10.699) | (5.802, 11.140) | (5.833, 11.191) |
| Secondary | 5.853*** | 5.876*** | 5.940*** |
|  | (4.337, 7.368) | (4.323, 7.429) | (4.366, 7.514) |
| Employed | 2.809*** | 2.754*** | 2.716*** |
|  | (1.594, 4.024) | (1.508, 4.000) | (1.465, 3.966) |
| Living arrangements |  |  |  |
| Living alone | -2.657 | -2.240 | -2.278 |
|  | (-6.200, 0.886) | (-5.910, 1.430) | (-5.963, 1.406) |
| Living in a couple | -2.902* | -2.478 | -2.516 |
|  | (-6.299, 0.494) | (-6.000, 1.045) | (-6.054, 1.023) |
| Living with relatives | -3.976** | -3.338* | -3.331* |
|  | (-7.374, -0.577) | (-6.867, 0.191) | (-6.881, 0.219) |
| Dependent children | -0.738 | -0.555 | -0.585 |
|  | (-2.145, 0.670) | (-1.991, 0.882) | (-2.027, 0.857) |
| Foreign born | -5.142*** | -4.971*** | -5.016*** |
|  | (-6.413, -3.872) | (-6.271, -3.670) | (-6.323, -3.709) |
| Use contraceptive methods | -0.764 | -0.246 | -0.277 |
|  | (-1.984, 0.456) | (-1.499, 1.007) | (-1.540, 0.986) |
| Publicly funded | -0.544 | 0.009 | 0.034 |
|  | (-2.299, 1.212) | (-1.900, 1.918) | (-1.884, 1.952) |
| Observations | 29,079 | 27,289 | 27,083 |
| R-squared | 0.150 | 0.150 | 0.147 |
| Adults only? | No | Yes | Yes |
| Elective only? | No | No | Yes |

Note: Reference category: women aged 40-44, with primary education or less, not employed, living with others, with no children in charge, born in Spain, not using contraceptive methods, whose abortion has not been publicly funded, and whose province of residence is Álava.
95% confidence intervals based on robust standard errors in parentheses. *** p<0.01, ** p<0.05, * p<0.1

The spacing between abortions decreases with a woman's age. For instance, the time interval for women aged 15-19 is 6 years less (-72.917 months, [-76.472, -69.361]) than among women aged 40-44 (reference: 0), and among women 35-39 is 1.4 years less (-16.705 months, [-20.282, -13.128]) than among women aged 40-44 (reference: 0). In terms of socioeconomic status, women with a university degree (8.123, [5.546, 10.699]) and those with secondary education (5.853, [4.337, 7.368]) exhibit a longer time interval than women with primary education or less (reference: 0). Employed women also take longer in between abortions (2.809, [1.594, 4.024]) than women who are not employed. When looking at living arrangements, it appears that women who live with their relatives have a shorter time interval (-3.976, [-7.374, -0.577]) than those who live with other people. We also find that foreign-born women (-5.142, [-6.413, -3.3872]) exhibit a shorter time interval between abortions than native women. No significant differences are documented in time intervals depending on whether women have dependent children (-0.738, [-2.145, 0.670]), whether women use contraceptive methods regularly (-0.764, [-1.984, 0.456]), or whether the abortion was publicly funded (-0.544, [-2.299, 1.121]).

While the qualitative and quantitative results in columns (2) and (3) are similar, the living arrangement finding does not seem to be consistent across columns. **S3 Table in S1 Appendix** contains regression estimates using the log of months, displaying results that are qualitatively the same.

## Discussion

Our findings on the likelihood of a repeat vs. first abortion indicate that it is particularly low education, having dependent children, living alone or being foreign-born that are associated with a higher likelihood of having a repeat abortion rather than a first abortion in 2019 in Spain. These findings resonate with previous studies, which have highlighted nationality and prior childbirth as significant correlates of repeat abortions in the Netherlands [9], and the role of immigrant status in the Basque Country, Spain [19]. While our findings are consistent with previous research [20], the high correlations between repeat abortion and both low education and being foreign-born are striking.

In examining the determinants of the actual number of abortions, we found that low education, not being employed, having dependent children, or being foreign-born served as important correlates. A study conducted in Ontario in 1998-1999 found that higher-order abortions were associated with being older, using contraception, being foreign-born, and having a history of physical and sexual violence by a male partner [21].

Our estimates further reveal that low education, not being employed, or foreign-born status are associated with a shorter interval between the last two abortions: low educated (6-8 months shorter interval), not being employed (3 months shorter interval), or foreign-born (5 months shorter interval). A unique study examining the timing to second abortion, focusing on women discharged from a New Zealand public hospital abortion clinic, found that younger age, non-European ethnicity, and number of children were associated with having a second abortion sooner [22]. A similar pattern regarding age and number of children was found in a region of Scotland, although there was no information on the socioeconomic characteristics of these women [23]. In a representative sample of women living in Britain during 2000-2001, it was measured that half of all second abortions reportedly occurred within 41 months of the previous procedure, and only 10% occurred more than 15 years apart; however, these findings were not linked to any health or socioeconomic characteristics of women [24].

## Conclusion

We conducted a comprehensive study on the correlates of repeat abortions and their spacing in Spain, using administrative data from the year preceding the COVID-19 pandemic. Despite the easy accessibility of contraceptive services in Spain, we found a 36% prevalence of repeat abortions, which is relatively high, but lower than in Canada or the US.

We find that compared to women undergoing their first abortion, those having repeat abortions are typically less educated, more likely to have dependent children, and more likely to live alone or be foreign-born. We also unveil a non-monotonic relationship with age. Notably, factors such as low education, not being employed, having dependent children, and foreign-born status correlate strongly with a higher number of repeat abortions, and a shorter time interval between them.

This article studies all induced abortions in a European country in a recent year making use of a comprehensive set of demographic and socioeconomic characteristics, including information on foreign-born status, and the month and year of both the current and previous abortions, if any. However, our study has two main limitations. Firstly, our estimated associations must be interpreted as correlations, not causal effects, given the cross-sectional observational data used. Secondly, information on previous abortions is self-reported, which can be problematic as women may find it difficult to admit to having had prior abortions [25]. Moreover, this reporting issue may vary with women's characteristics.

Despite these limitations, our findings contribute to a better understanding of repeat abortions. We strongly recommend healthcare services and policymakers to target family planning and contraception services more effectively following a woman's first abortion. Special attention should be paid to high-risk groups. In Spain, those high-risk groups include the low-educated, not employed, and foreign-born.

# Acknowledgments
We thank Carlos Guillermo Bozzoli, Damian Clarke, Pradeep Kumar and Ines Lee for comments and suggestions.

# Funding
Catia Nicodemo acknowledges funding from the Economic and Social Research Council [grant number ES/T008415/1] and from the National Institute for Health Research Applied Research Collaboration Oxford and Thames Valley at Oxford Health NHS Foundation Trust.

# Author contributions
**Conceptualization:** Catia Nicodemo, Sonia Oreffice, Climent Quintana-Domeque.

**Data curation:** Catia Nicodemo, Sonia Oreffice, Climent Quintana-Domeque.

**Formal analysis:** Catia Nicodemo, Sonia Oreffice, Climent Quintana-Domeque.

**Investigation:** Catia Nicodemo, Sonia Oreffice, Climent Quintana-Domeque.

**Methodology:** Catia Nicodemo, Sonia Oreffice, Climent Quintana-Domeque.

**Writing – original draft:** Catia Nicodemo, Sonia Oreffice, Climent Quintana-Domeque.

**Writing – review & editing:** Catia Nicodemo, Sonia Oreffice, Climent Quintana-Domeque.

# S1 Appendix

**S1 Table. Correlates of repeat abortion among women who had an abortion in 2019.**
**Dependent variable = 1 if repeat abortion, = 0 no previous abortion**
**Logit probability model: Odds ratios estimates (95% Confidence intervals)**

|  | (1) | (2) | (3) |
|---|---|---|---|
| Age |  |  |  |
| 15-19 | 0.318*** | 0.403*** | 0.386*** |
|  | (0.290 , 0.348) | (0.366 , 0.444) | (0.350 , 0.427) |
| 20-24 | 0.869*** | 0.860*** | 0.819*** |
|  | (0.813 , 0.930) | (0.804 , 0.919) | (0.764 , 0.879) |
| 25-29 | 1.189*** | 1.182*** | 1.134*** |
|  | (1.117 , 1.266) | (1.110 , 1.259) | (1.061 , 1.211) |
| 30-34 | 1.250*** | 1.247*** | 1.217*** |
|  | (1.175 , 1.330) | (1.172 , 1.326) | (1.140 , 1.299) |
| 35-39 | 1.115*** | 1.114*** | 1.096*** |
|  | (1.047 , 1.188) | (1.046 , 1.187) | (1.025 , 1.172) |
| Education |  |  |  |
| University | 0.361*** | 0.355*** | 0.364*** |
|  | (0.340 , 0.383) | (0.335 , 0.377) | (0.342 , 0.387) |
| Secondary | 0.787*** | 0.775*** | 0.776*** |
|  | (0.753 , 0.822) | (0.741 , 0.810) | (0.741 , 0.813) |
| Employed | 0.937*** | 0.925*** | 0.938*** |
|  | (0.907 , 0.968) | (0.895 , 0.956) | (0.906 , 0.970) |
| Living arrangements |  |  |  |
| Living alone | 1.479*** | 1.484*** | 1.489*** |
|  | (1.353 , 1.616) | (1.357 , 1.622) | (1.357 , 1.633) |
| Living in a couple | 1.268*** | 1.270*** | 1.330*** |
|  | (1.165 , 1.381) | (1.166 , 1.384) | (1.217 , 1.453) |
| Living with relatives | 1.238*** | 1.256*** | 1.268*** |
|  | (1.134 , 1.351) | (1.150 , 1.371) | (1.158 , 1.389) |
| Dependent children | 1.858*** | 1.845*** | 1.824*** |
|  | (1.791 , 1.927) | (1.779 , 1.914) | (1.756 , 1.895) |
| Foreign born | 1.240*** | 1.233*** | 1.209*** |
|  | (1.200 , 1.282) | (1.193 , 1.275) | (1.168 , 1.251) |
| Use contraceptive methods | 1.082*** | 1.085*** | 1.035** |
|  | (1.048 , 1.116) | (1.051 , 1.120) | (1.001 , 1.070) |
| Publicly funded | 1.098*** | 1.093*** | 1.126*** |
|  | (1.054 , 1.144) | (1.049 , 1.139) | (1.076 , 1.177) |
| Observations | 81,168 | 78,129 | 72,052 |
| Adults only? | No | Yes | Yes |
| Elective only? | No | No | Yes |

Note: Reference category: women aged 40-44, with primary education or less, not employed, living with others, with no children in charge, born in Spain, not using contraceptive methods, whose abortion has not been publicly funded, and whose province of residence is Álava.
95% confidence intervals in parentheses.
*** p<0.01, ** p<0.05, * p<0.1

**S2 Table. Correlates of number of previous abortions among women who had an abortion in 2019.**
**Dependent variable = previous number of abortions**
**Poisson regression: IRR estimates (95% Confidence intervals)**

|  | (1) | (2) | (3) |
|---|---|---|---|
| Age |  |  |  |
| 15-19 | 0.329*** | 0.415*** | 0.400*** |
|  | (0.305, 0.354) | (0.384, 0.449) | (0.369, 0.433) |
| 20-24 | 0.841*** | 0.836*** | 0.809*** |
|  | (0.801, 0.883) | (0.796, 0.877) | (0.769, 0.851) |
| 25-29 | 1.118*** | 1.115*** | 1.080*** |
|  | (1.069, 1.169) | (1.066, 1.165) | (1.031, 1.130) |
| 30-34 | 1.170*** | 1.168*** | 1.145*** |
|  | (1.120, 1.221) | (1.118, 1.220) | (1.095, 1.198) |
| 35-39 | 1.078*** | 1.077*** | 1.061** |
|  | (1.031, 1.127) | (1.030, 1.126) | (1.013, 1.111) |
| Education |  |  |  |
| University | 0.409*** | 0.407*** | 0.417*** |
|  | (0.391, 0.428) | (0.389, 0.425) | (0.398, 0.436) |
| Secondary | 0.779*** | 0.774*** | 0.776*** |
|  | (0.757, 0.801) | (0.752, 0.797) | (0.753, 0.799) |
| Employed | 0.906*** | 0.900*** | 0.909*** |
|  | (0.885, 0.927) | (0.880, 0.921) | (0.888, 0.931) |
| Living arrangements |  |  |  |
| Living alone | 1.318*** | 1.320*** | 1.327*** |
|  | (1.230, 1.413) | (1.232, 1.415) | (1.235, 1.425) |
| Living in a couple | 1.184*** | 1.185*** | 1.223*** |
|  | (1.107, 1.267) | (1.108, 1.268) | (1.140, 1.311) |
| Living with relatives | 1.167*** | 1.176*** | 1.187*** |
|  | (1.089, 1.251) | (1.098, 1.261) | (1.105, 1.275) |
| Dependent children | 1.613*** | 1.605*** | 1.585*** |
|  | (1.568, 1.659) | (1.561, 1.651) | (1.540, 1.632) |
| Foreign born | 1.132*** | 1.129*** | 1.110*** |
|  | (1.106, 1.159) | (1.102, 1.155) | (1.083, 1.137) |
| Use contraceptive methods | 0.990 | 0.991 | 0.961*** |
|  | (0.969, 1.012) | (0.969, 1.013) | (0.939, 0.983) |
| Publicly funded | 1.061*** | 1.058*** | 1.074*** |
|  | (1.029, 1.093) | (1.027, 1.091) | (1.040 - 1.110) |
| Observations | 80,800 | 77,761 | 71,709 |
| Adults only? | No | Yes | Yes |
| Elective only? | No | No | Yes |

Note: Reference category: women aged 40-44, with primary education or less, not employed, living with others, with no children in charge, born in Spain, not using contraceptive methods, whose abortion has not been publicly funded, and whose province of residence is Álava.

95% confidence intervals based on robust standard errors in parentheses.

*** $p<0.01$, ** $p<0.05$, * $p<0.1$

**S3 Table. Correlates of abortion spacing among women who had an abortion in 2019.**
**Dependent variable = log(months between current abortion and previous abortion)**
**Linear regression model: OLS regression estimates (95% Confidence intervals)**

|  | (1) | (2) | (3) |
|---|---|---|---|
| Age |  |  |  |
| 15-19 | -1.426*** | -1.439*** | -1.403*** |
|  | (-1.496 , -1.357) | (-1.511 , -1.366) | (-1.479 , -1.327) |
| 20-24 | -1.027*** | -1.031*** | -1.033*** |
|  | (-1.079 , -0.975) | (-1.085 , -0.977) | (-1.087 , -0.979) |
| 25-29 | -0.650*** | -0.656*** | -0.657*** |
|  | (-0.699 , -0.600) | (-0.707 , -0.604) | (-0.708 , -0.605) |
| 30-34 | -0.398*** | -0.403*** | -0.403*** |
|  | (-0.448 , -0.349) | (-0.454 , -0.351) | (-0.455 , -0.352) |
| 35-39 | -0.222*** | -0.219*** | -0.220*** |
|  | (-0.273 , -0.171) | (-0.273 , -0.166) | (-0.273 , -0.166) |
| Education |  |  |  |
| University | 0.119*** | 0.124*** | 0.124*** |
|  | (0.070 , 0.168) | (0.073 , 0.174) | (0.074 , 0.175) |
| Secondary | 0.102*** | 0.103*** | 0.104*** |
|  | (0.071 , 0.134) | (0.070 , 0.135) | (0.071 , 0.137) |
| Employed | 0.060*** | 0.059*** | 0.058*** |
|  | (0.035 , 0.085) | (0.034 , 0.085) | (0.032 , 0.083) |
| Living arrangements |  |  |  |
| Living alone | -0.057 | -0.049 | -0.049 |
|  | (-0.126 , 0.013) | (-0.121 , 0.024) | (-0.122 , 0.023) |
| Living in a couple | -0.047 | -0.038 | -0.039 |
|  | (-0.114 , 0.020) | (-0.108 , 0.032) | (-0.109 , 0.031) |
| Living with relatives | -0.057* | -0.047 | -0.046 |
|  | (-0.126 , 0.011) | (-0.119 , 0.024) | (-0.118 , 0.026) |
| Dependent children | -0.015 | -0.017 | -0.018 |
|  | (-0.043 , 0.014) | (-0.046 , 0.013) | (-0.048 , 0.011) |
| Foreign born | -0.065*** | -0.064*** | -0.066*** |
|  | (-0.090 , -0.039) | (-0.090 , -0.038) | (-0.092 , -0.040) |
| Use contraceptive methods | -0.004 | 0.005 | 0.003 |
|  | (-0.028 , 0.020) | (-0.020 , 0.030) | (-0.022 , 0.028) |
| Publicly funded | 0.011 | 0.030 | 0.031* |
|  | (-0.022 , 0.045) | (-0.006 , 0.067) | (-0.006 , 0.068) |
| Observations | 29,079 | 27,289 | 27,083 |
| R-squared | 0.127 | 0.127 | 0.122 |
| Adults only? | No | Yes | Yes |
| Elective only? | No | No | Yes |

Note: Reference category: women aged 40-44, with primary education or less, not employed, living with others, with no children in charge, born in Spain, not using contraceptive methods, whose abortion has not been publicly funded, and whose province of residence is Álava.
95% confidence intervals based on robust standard errors in parentheses.
*** $p<0.01$, ** $p<0.05$, * $p<0.1$